\documentclass[global,twocolumn]{svjour}
\usepackage{graphics}
\journalname{Appl. Phys. B}
\begin{document}
\title{A laser setup for rubidium cooling dedicated to space applications}
\author{T.~L\'{e}v\`{e}que \and L.~Antoni-Micollier\thanks{\emph{Present address:}
LP2N - Institut d'Optique d'Aquitaine, All\'{e}e Laroumagne, 33400 Talence, France} \and B.~Faure \and J.~Berthon \mail{thomas.leveque@cnes.fr}}
\institute{Centre National d'\'{E}tudes Spatiales, 18 avenue Edouard Belin, 31400 Toulouse, France}
\date{Received: date / Revised version: date}
\maketitle
\begin{abstract}
We present the complete characterization of a laser setup for rubidium cooling dedicated to space applications. The experimental setup is realized with commercial off-the-shelf fiber components suitable for space applications. By frequency doubling two fiber laser diodes at 1560~nm, we produce the two optical frequencies at 780~nm required for atomic cooling of $^{87}$Rb. The first laser is locked on saturated absorption signal and long term frequency drift has been canceled using a digital integrator. The optical frequency of the second laser is controlled relatively to the first one by a frequency comparison method. A full characterization of the setup, including frequency stability evaluation and frequency noise measurement has been performed. The optical frequency doubling module has been submitted to environmental tests to verify its compatibility with space applications.
\end{abstract}

\section{Introduction}

Over the past decades, the development of laser techniques for cooling and manipulating atoms has contributed to the advent of a new generation of atom-based instruments~\cite{sortais2001,peters2001}. Cold atom sensors have become the state-of-the-art for metrological applications~\cite{gauguet2009}. These instruments are as useful for fundamental physics measurements~\cite{hogan2011,clade2006,wolf2006,geiger2011} as for applied metrology~\cite{bodart2009,bidel2013}. Besides, utilization of cold atom sensors in space is promising as they benefit from the zero gravity conditions which enable to interact with the atomic system during a longer time, increasing the performances of the device. For instance, the PHARAO space clock, part of the ACES space project~\cite{cacciapuoti2007}, will run with a potential accuracy of $10^{-16}$ by interacting with the atoms up to 2~s. Dual-species cold atom interferometers are also considered in the frame of the STE-QUEST space mission~\cite{cacciapuoti2012} in order to realize a test of the equivalence principle at the level of $10^{-15}$. Cold atom gradiometers could also be considered for next generation of geodesic space sensors.

Rubidium-based cold atom apparatus, which have been extensively studied, are promising in the frame of space applications. Cold atomic clouds prepared with sub-Doppler cooling methods can be used directly by the instrument, or can be a step to Bose-Einstein condensation with further cooling process~\cite{muntinga2013}. Developing a standard and space qualified optical setup to fulfill this function is then necessary for the realization of any cold atom experiments in space.

Development of space instruments is technically challenging. These systems must be reliable enough to undergo high level of vibration and radiation and must remain operational over a large range of temperature. The laser source is one of the most critical part of a cold atom apparatus. It must keep its performances in terms of delivered optical power, linewidth, frequency stability and spectral agility in a space environment. The system must work autonomously over the entire mission lifespan, which is typically of 24 months, without external tuning. The only laser cooling setup developed so far for the PHARAO space project is based on free space external-cavity diode lasers~\cite{baillard2006,schmidt2011} (ECDL) amplified by injection in slave laser diodes. These custom-made free space architectures are very sensitive to mechanical misalignment caused by vibration or thermal variations. Alternative technical approach, based on fiber telecom technologies, has been recently investigated to alleviate these problems~\cite{stern2010,menoret2011,sane2012}. The Rubidium transitions at $\lambda$=780~nm can be addressed via a frequency doubling process of 1560~nm telecom C-band laser sources.

In this paper, we report on the characterization of a laser source, dedicated to rubidium cooling, based on an assembly of commercial off-the-shelf telecom fiber components. The optical setup is presented in Sect.~\ref{optical_architecture}. The level of space qualification of each key components has been evaluated. Especially, the frequency doubling module has been submitted to environmental tests summarized in Sect.~\ref{trl}. A complete operational setup has been realized. Sections~\ref{rep} and \ref{ref} are dedicated to frequency lock techniques and spectral characterization of the lasers.

\section{Optical setup} \label{optical_architecture}

\begin{figure}[!h]
\centering \resizebox{8cm}{!}{\includegraphics{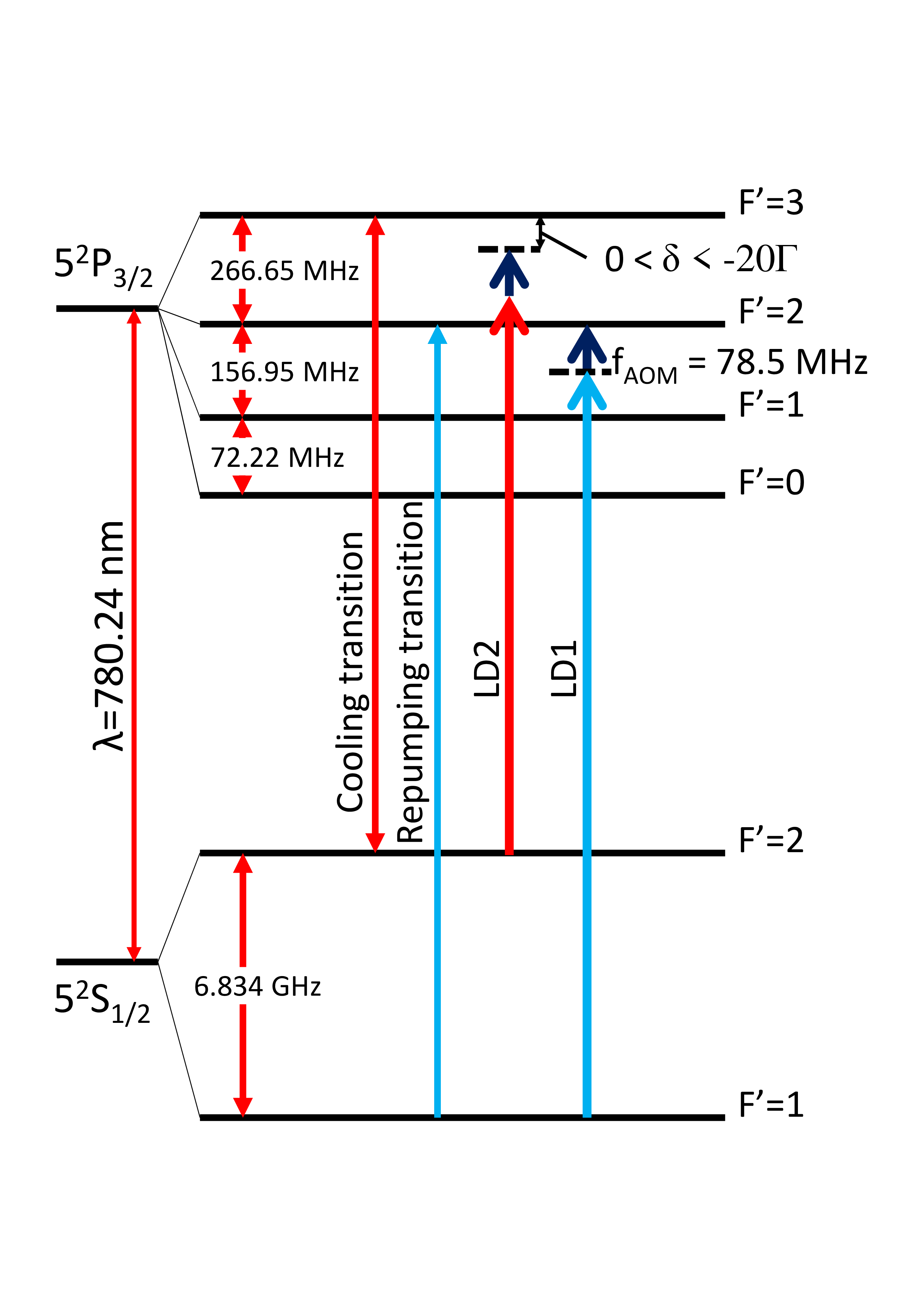}} 
\caption{$^{87}$Rb D2 transition hyperfine structure. LD1 and LD2 are respectively used for driving the repumping and cooling transitions.}
\label{niveaux_rb}
\end{figure}

Two optical frequencies are needed to perform the laser cooling of $^{87}$Rb. The first one, tuned around the $|5S_{1/2},$ $F = 2 \rangle$ $\rightarrow$ $|5P_{3/2},F' = 3 \rangle$ cycling transition, is used to drive the cooling process. The second one, tuned on the $|5S_{1/2},F = 1 \rangle$ $\rightarrow$ $|5P_{3/2},F' = 2 \rangle$ transition, repumps the atoms from the $|5S_{1/2},F = 1 \rangle$ ground state back into the cooling cycle. The frequency difference between these two transitions is of 6.568~GHz. The $^{87}$Rb D2 transition hyperfine structure~\cite{steck2010} is reported in Fig.~\ref{niveaux_rb}.

\begin{figure}[!h]
\centering \resizebox{8cm}{!}{\includegraphics{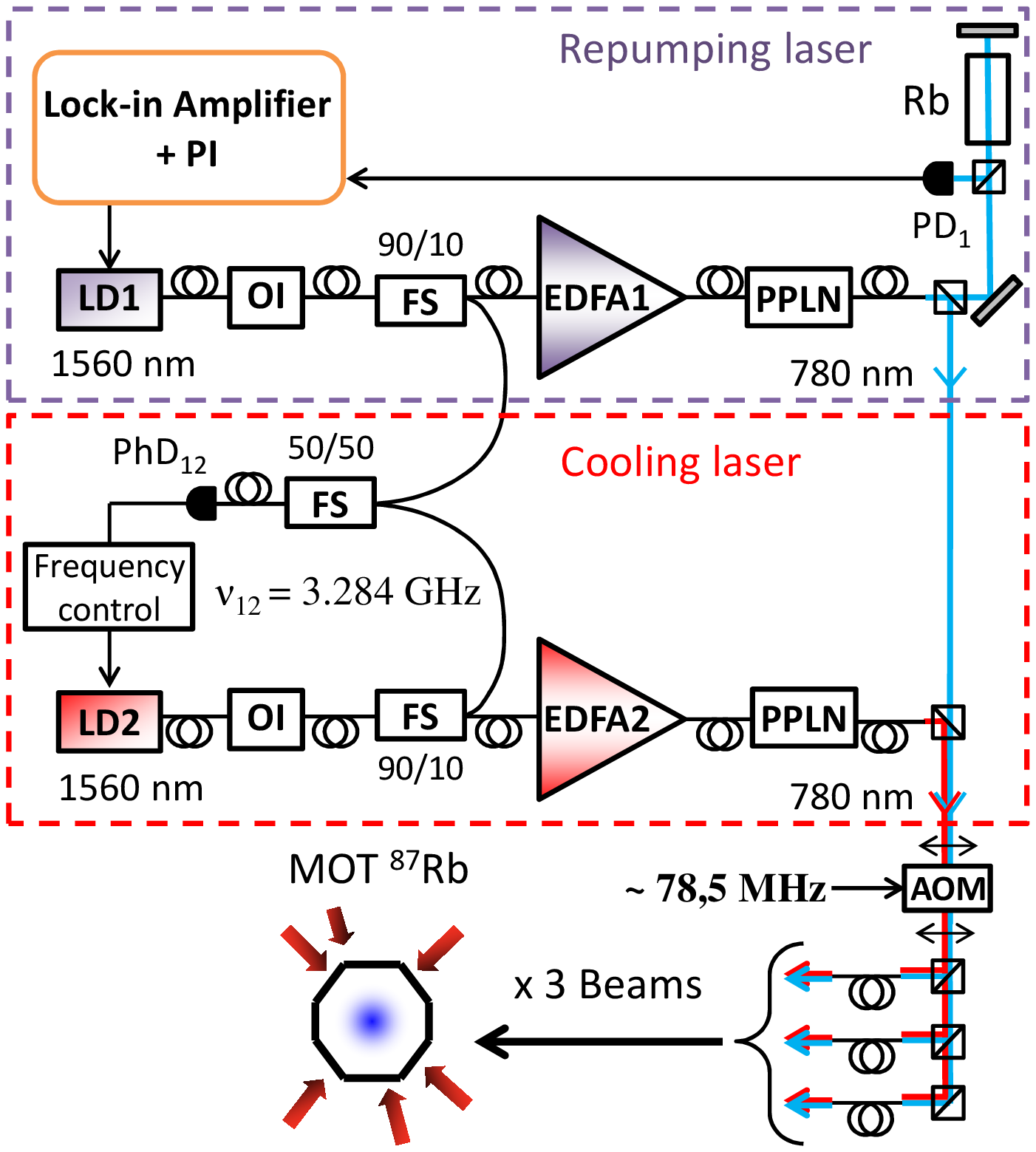}} 
\caption{Diagram of the complete laser system. LD1 and LD2: RIO laser diodes at 1560~nm; OI: optical isolator; PPLN: periodically poled lithium niobate crystal; AOM: acousto-optic modulator; FS: fiber splitter; MOT: magneto-optical trap.}
\label{setup}
\end{figure}

The laser setup, depicted in Fig.~\ref{setup}, consists of two monoblock butterfly packaged ECDL (LD1 and LD2, RIO Planex) emitting at 1560~nm,  amplified with two Yb-Er doped amplifiers~\cite{ampliEDFA} and doubled in frequency through fiber periodically poled Lithium Niobate (PPLN) waveguides (WH-0780-000-F-B-C, NTT Electronics). These ensembles form two fiber laser sources at 780~nm necessary to generate the repumping and cooling laser frequencies for laser cooling.

The ECDL  are based on an intracavity wavelength selection by a planar Bragg reflector waveguide~\cite{numata2010} and benefit from a narrow linewidth (5~kHz) and a wide enough tunability ($\sim$10~GHz). The diodes are regulated around room temperature and supplied by a current of $\sim$90~mA to provide an optical output power of 15~mW. The frequency control is achieved by a feedback on the diode current.

The first laser LD1 is used, after the PPLN doubling stage, to drive the repumping transition. It is frequency locked by saturation spectroscopy on the prominent cross-over signal between the $|5P_{3/2},F' = 1 \rangle$ and $|5P_{3/2}, F' = 2 \rangle$ atomic states, red-detuned by 78.5~MHz from the repumping transition. This laser is then used as an absolute frequency reference to control the frequency of LD2 which drives the cooling transition after the PPLN doubling stage. The frequency difference between LD1 and LD2 at 1560~nm is set around $\nu _{12}$ = 3.284~GHz by the method described in Sect.~\ref{ref}.

In order to provide sufficient optical power, the output signals of LD1 and LD2 are injected in two fiber amplifiers. In normal operation, $\sim$150~mW of LD1 and $\sim$360~mW of LD2 are obtained after amplification. After the frequency doubling stage, the frequency difference between the two lasers is of 6.568~GHz, corresponding to the frequency difference between the repumping and cooling transitions. In this configuration, we obtain $\sim$45~mW and $\sim$180~mW of repumping and cooling light respectively, at the output of the doubling stages.

Then the output beams are combined and frequency shifted by an acousto-optic modulator (AOM, MT80-B301.5-IR, A$\&$A), from which the first output order is separated and coupled to three polarization maintaining fibers. The AOM is driven at $f_{\mathrm{AOM}}$=78.5~MHz in order to bridge the gap between the repumping transition and the cross-over signal on which LD1 is locked. This scheme enables the laser light to be intensity controlled by changing the input RF power of the AOM. The three fiber output beams, containing $\sim$30~mW of cooling laser and $\sim$4~mW of repumping laser, can be used for realizing a magneto-optical trap in a retro-reflected configuration.

\section{Compatibility with space environment} \label{trl}

In this part, we report on the compatibility of the optical setup, described in Sect.~\ref{setup}, with space environment. In order to facilitate the qualification process, we realized the optical setup with commercial Telcordia-qualified components~\cite{telcordia2004}. Specifications of reliability of these components are close to those required for space applications. Radiation is the only main field which is not covered by the Telcordia standard. Few efforts are then required for making them compatible with space applications.

The Telecordia-qualified planar waveguide ECDL, built by Redfern Integrated Optics, offers advantages over free-space ECDL. Its simpler design allows a better stability and robustness of the device. The component is butterfly packaged and benefits from a more compact size and lower mass compared to other ECDL designs. Complementary qualification studies, including radiation tests, are carried out to ensure a full compatibility with space applications~\cite{numata2012}.

Most of Er/Yb doped optical fiber amplifiers are now Telcordia qualified. Nevertheless previous studies showed that doped optical fibers are the most radiation sensitive part of amplifiers. They show a high radiation sensitivity compared to passive optical fibers when exposed to X-rays, gamma-rays or protons. As a consequence, the estimation of their vulnerability to the hard environments associated with space missions remains crucial. Several studies~\cite{girard2012,thomas2012} have shown that combining different hardening approaches, it appears possible to design radiation-hardened fibers for doses up to 150~krad, compatible with orbital space missions in the near future.

The frequency doubling module is the keystone of the optical setup. The compact packaging of the component (32 x 12 x 54~mm) includes a pigtailed PPLN waveguide and an internal thermo-electric cooler which enables to control its temperature. This component has no specific space qualification and its performances are guaranteed by the manufacturer for a maximum input power of 200~mW. In order to evaluate its compatibility with space applications, we performed a complete set of qualification tests on four devices.

\begin{figure}[!h]
\centering \resizebox{8cm}{!}{\includegraphics{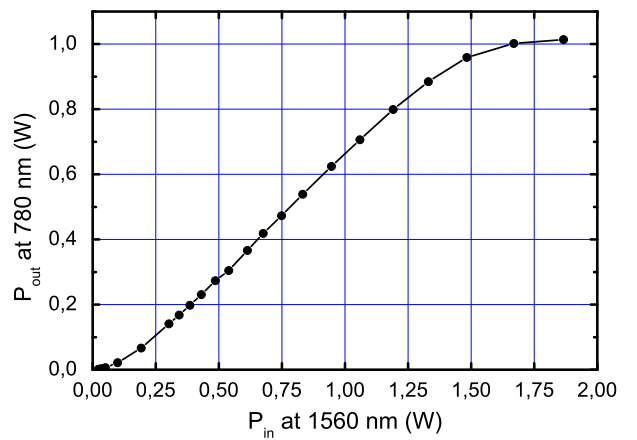}} 
 \caption{Evolution of 780~nm output power of the frequency conversion module $P_{\mathrm{out}}$ as a function of the input power $P_{\mathrm{in}}$ at 1560~nm. The temperature of the component is set at $T_{\mathrm{pm}}$=65$^{\circ}$C.}
 \label{doubleur}
 \end{figure}

First, we characterize the output power of the component at 780~nm as a function of the input power at 1560~nm. The measurement of the output power is given in Fig.~\ref{doubleur} and shows a maximum of 1~W for an input power of 1.8~W at 1560~nm. This device is not submitted to the environmental tests and is used as a reference to verify and prevent any drift of the measurement scheme.

The conversion efficiency $\eta$, defined as:

\begin{equation}
\eta(\%/\mathrm{W}) = \frac{P_{\mathrm{out}}(\mathrm{W})}{P_{\mathrm{in}}(\mathrm{W})^2}\times 100
\end{equation}

is around 250$\%$/W for an input power of 20~mW. It is maximal when the device is regulated at the optimal phase-matching temperature $T_{\mathrm{pm}}$ which is typically between 40$^{\circ}$C and 65$^{\circ}$C. The parameters $\eta$ and $T_{\mathrm{pm}}$ can be experimentally determined with a respective relative uncertainty of 13$\%$ and 5$\%$ at 3$\sigma$. These two parameters are measured before and after each environmental test to monitor any functional evolution of the components.

First, we perform a lifetime test on three components at different levels of input power. The input power of each component is respectively set at 400~mW, 700~mW and 1000~mW during 2000~hours. Despite a higher input power than specified by the manufacturer, no measurable evolution of the parameters $\eta$ and $T_{\mathrm{pm}}$ were observed on these three devices after the life test, within the measurement uncertainty (see Appendix \ref{environmental_tests}). Then, one of these components was submitted to a set of environmental tests representative of space conditions related to  standard orbital missions. Irradiation was carried out for a dose up to 50~krad at 360 rad/h. After, the device was submitted to 50 thermal cycles between -40$^{\circ}$C and +60$^{\circ}$C (slope of 1$^{\circ}$C/min) and then to 50 additional cycles  between -55$^{\circ}$C and +75$^{\circ}$C. A random vibration test has been realized between 50~Hz and 2~kHz with an overall level of 20~g$_{RMS}$~\cite{mil}. Finally, a shock test has been performed at a level of 500~g. No measurable evolution of $\eta$ and $T_{\mathrm{pm}}$ were observed during these tests within the measurement uncertainty (see Appendix \ref{environmental_tests}). The results of this preliminary evaluation show that the component would be suitable for space application. Identical complementary tests could be performed on several components to get a more statistically representative qualification.

The free-space part of the setup including AOM, power splitting and injection in the fiber, has been qualified in the frame of the PHARAO space project. This solution remains more stable in terms of power and polarization than commercial fiber components at 780~nm. 

\section{Repumping laser}\label{rep}

The first laser LD1 is frequency doubled, to be resonant with the repumping transition during the cooling process.
It is locked on a saturated absorption signal of $^{87}$Rb using a frequency modulation spectroscopy technique~\cite{hall1981}. The current of the laser diode is modulated at a frequency of 100~kHz. After the doubling stage, the laser light at $\lambda$=780~nm is sent to a rubidium cell in a retro-reflected configuration to realize a saturated absorption spectroscopy. The signal is collected on a photodiode (PD$_{1}$, PDA36A-EC, Thorlabs, 10~MHz bandwidth) and demodulated with a lock-in amplifier (SR830, Stanford Research Systems). The error signal is processed to generate the correction signal applied to the laser current. The temperature of the laser diode is regulated at 31$^{\circ}$C through the internal temperature control.

The current-temperature couple setpoint of the laser, for a given output optical frequency, is drifting over time. This drift is thought to be due to residual thermal effects in the Bragg mirror waveguide inside the butterfly module. In order to guarantee the stability of the laser frequency on the short and long term, two types of corrections are generated from the error signal. The first one is made by a standard analog proportional-integral corrector which controls the laser frequency in a bandwidth of 1.2~kHz. Moreover, we set up a digital integrator to cancel the long-term drift of the analog laser lock. The error signal $\epsilon(t)$ is measured and averaged over a typical time of $\tau_c$=50~s by a data acquisition module (NI USB-6259). An incremental cyclic correction $C_{i}$, given by the equation~\ref{integrateur}, is then calculated and added to the analog correction signal every 50~ms. The gain parameter $\alpha$ is adjusted to reach the best stability on the error signal. This correction ensures a robust frequency stability on long time scale without external thermal regulation of the laser case or active compensation of the temperature setpoint.

\begin{equation}
\label{integrateur}
C_{i+1}=\sum_{j=1}^{i}{C_{j}} + \alpha \int_{t_i}^{t_i+\tau_c}{\epsilon(t) dt}
\end{equation}

\begin{figure}[!h]
\centering \resizebox{8cm}{!}{\includegraphics{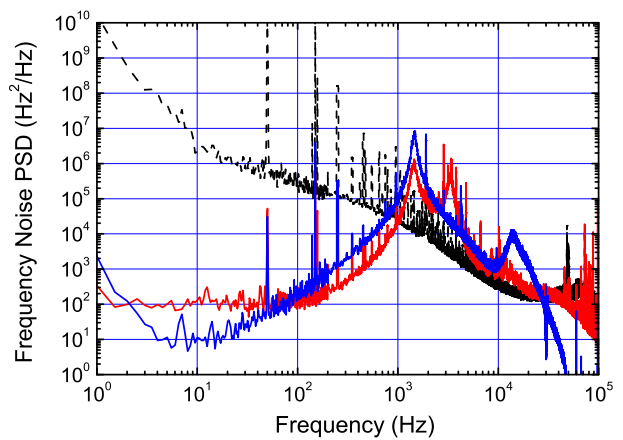}}
 \caption{(Color online) Frequency-noise power spectral density. The blue curve shows the in-loop noise of LD1. The red curve shows the in-loop noise of LD2. The dash black curve shows the free-running frequency noise spectrum measured on the beat note between LD1 and LD2.}
 \label{dsp}
 \end{figure}

In order to analyze the laser frequency noise, the in-loop power spectral density of the frequency noise is directly measured by recording the error signal given by the lock-in amplifier. The frequency range up to 100 kHz is measured with a data acquisition module providing a fast Fourier transformation routine (FFT). The frequency noise spectral density after the lock-in amplifier is displayed in Fig.~\ref{dsp}. It exhibits a maximum for a frequency of 1.2~kHz corresponding to the servo loop bandwidth. Above this frequency, the noise follows the typical free-running noise of the laser. The second maximum at 12~kHz originates from noise and internal filters of the lock-in amplifier.

\begin{figure}[!h]
\centering \resizebox{8cm}{!}{\includegraphics{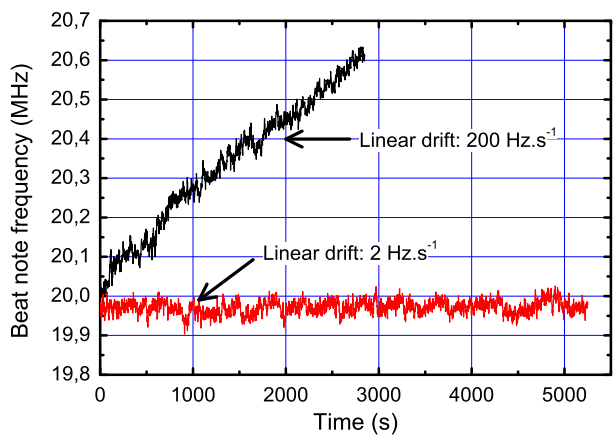}}
 \caption{Beat note frequency between LD1 and the OFC. The black curve shows the beat note frequency when LD1 is locked with the analog feed-back loop only. The red curve shows the beat note frequency when LD1 is locked thanks to the analog and digital feed-back loop.}
 \label{stability_rep}
 \end{figure}

\begin{figure}[!h]
\centering \resizebox{8cm}{!}{\includegraphics{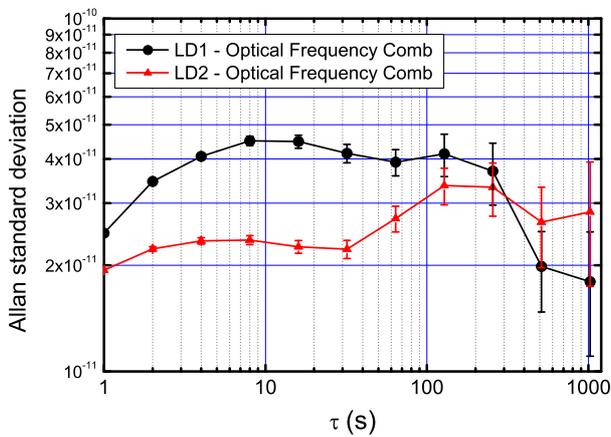}}
 \caption{Fractional frequency stability (Allan deviation)
Black circles: Typical stability of the beat note between LD1 with frequency stabilization and the OFC. Red triangles: Typical stability of the beat-note between LD2 with frequency stabilization and the OFC.}
 \label{allan_dev}
 \end{figure}

The frequency stability is measured by comparison with an Optical Frequency Comb~\cite{udem2002} (OFC, FC1500-250-WG, Menlo Systems). The OFC is referenced on an atomic clock~\cite{pajot2012} (OSA 3210) which provides a long term relative frequency stability around $3 \times 10^{-11}~\tau^{-1/2}$. The frequency of the beat note between LD1 and the OFC, around 20~MHz, is measured with a zero dead-time counter (FXM50). The measurement is displayed in Fig.~\ref{stability_rep}. First, we perform the measurement by locking the laser only with the analog feed-back loop. The signal exhibits a linear drift of 200~Hz.s$^{-1}$. Then a measurement is realized by locking the laser with the analog and digital feed-back loop. The linear drift is then reduced to 2~Hz.s$^{-1}$ over 5000~s. The Allan standard deviation of this sample is reported on Fig.~\ref{allan_dev}. It shows a stability under $5 \times 10^{-11}$ from 1~s to 1000~s.

\section{Cooling laser}\label{ref}

\begin{figure}[!h]
\centering \resizebox{8cm}{!}{\includegraphics{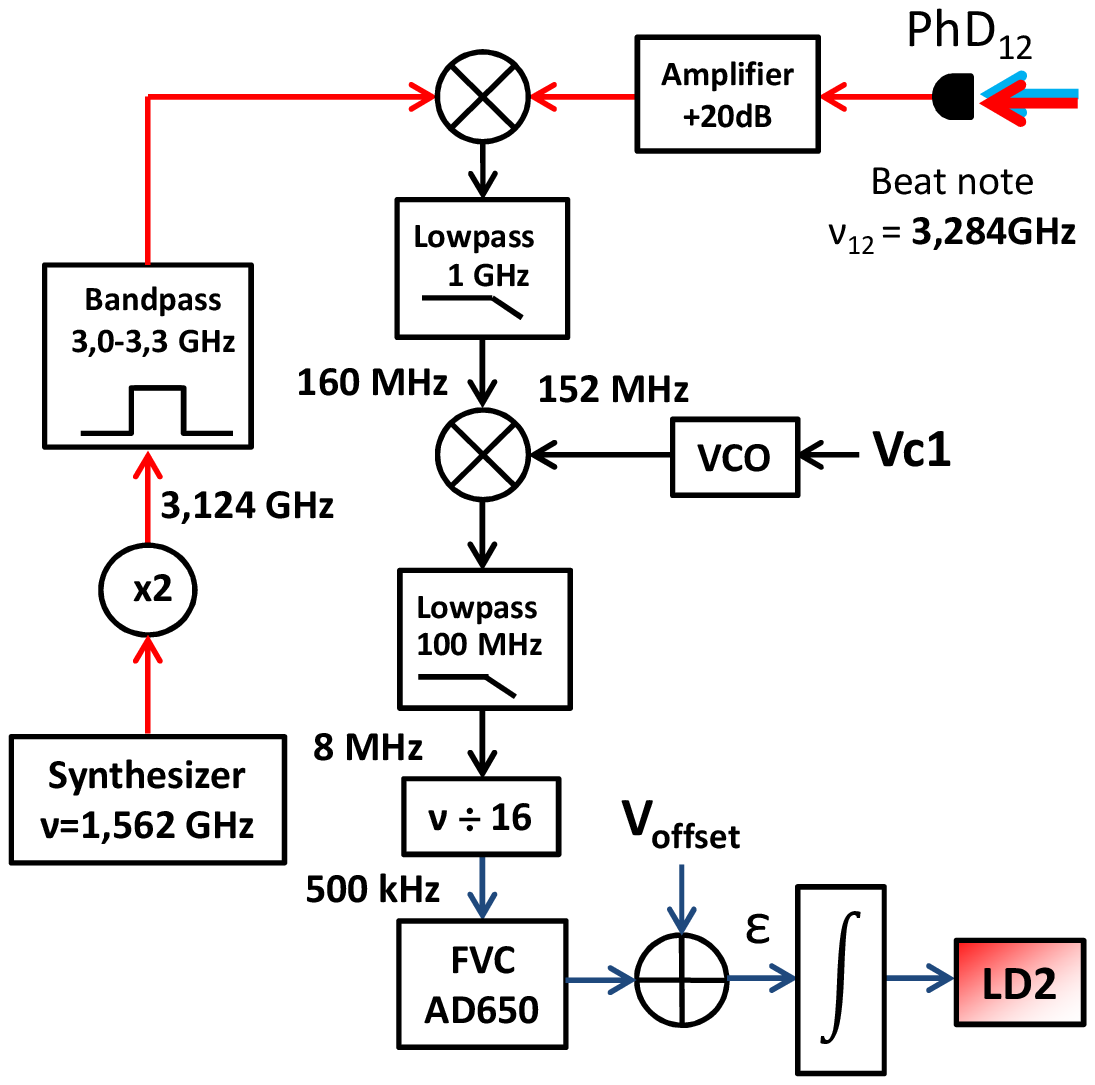}}
 \caption{Frequency conversion chain. The beat note between LD1 and LD2 is first mixed with a reference microwave signal at 3.124~GHz. The resulting frequency at 160~MHz is mixed with a VCO signal at 152~MHz. The output frequency is divided by 16 and converted to a tension through a FVC. This signal is offset and integrated to control the frequency of LD2.}
 \label{frequency_chain}
 \end{figure}

The second laser LD2 is used, after the frequency doubling stage, to drive the cooling transition $|5S_{1/2},F=2\rangle$ $\rightarrow$ $|5P_{3/2},F'=3\rangle$. The frequency difference between the repumping and cooling transitions is of 6.568~GHz~\cite{steck2010}. In order to perform sub-Doppler cooling, the detuning between the laser frequency and the cooling transition must be adjustable between 0 and -20~$\Gamma$ ($\Gamma$=6.06~MHz, natural linewidth of the $^{87}$Rb D2 line) on a typical duration of 1~ms. To fulfill these requirements, LD1 is used as a frequency reference. The frequency difference between LD2 and LD1 is then locked and controlled via a microwave system described in Fig.~\ref{frequency_chain}. Small amounts of light (10~$\%$) of LD1 and LD2 are superimposed on a fast photoconductor (PhD$_{12}$, Hamamatsu G7096-03, bandwidth: 15~GHz). The beat note at $\nu _{12}$~=~3.284~GHz is then amplified and mixed with a reference signal, given by a microwave synthesizer at $\nu$~=~1.562~GHz doubled in frequency. The output signal at $\nu_{\mathrm{int}}$~=~160~MHz is multiplied by the signal given by a Voltage Control Oscillator (VCO, DRFA10Y) at $\nu_{\mathrm{VCO}}$~=~152~MHz. The resulting signal frequency at 8~MHz is divided by 16 and sent to a Frequency-to-Voltage Converter (FVC, AD650) which derives an error signal proportional to the frequency difference between the two lasers. This signal is offset by $V_{\mathrm{offset}}$ in order to set the signal at zero when the frequency difference between $\nu_{\mathrm{VCO}}$ and $\nu_{\mathrm{int}}$ is equal to 8~MHz. After shaping and filtering, this output signal is used to generate the feedback allowing to frequency lock the laser LD2 on LD1.

The frequency $\nu_{\mathrm{VCO}}$ can be tuned from 152~MHz to 213~MHz by the signal Vc1 (33~MHz/V). When the servo loop is closed, this signal allows to drive the frequency difference between LD1 and LD2 over a span of 61~MHz at 1560~nm, which corresponds to a detuning of -20~$\Gamma$=122~MHz at 780~nm.

In order to analyze the laser frequency noise, the in-loop power spectral density of the frequency noise is directly measured by recording the error signal of the FVC. The frequency range up to 100 kHz is measured using a FFT routine and shown in Fig.~\ref{dsp}. It exhibits a maximum for a frequency of 3~kHz corresponding to the servo loop bandwidth. Above this frequency, the noise follows the typical free-running noise of the laser. The free-running frequency noise of the lasers is measured by unlocking the two lasers and recording the spectrum at the output of the FVC. The result, reported in Fig.~\ref{dsp} shows a 1/f dependence above 10~Hz. When the lasers are locked, the noise was suppressed by a factor of $\sim 10^6$ within the control bandwidth.

\begin{figure}[!h]
\centering \resizebox{8cm}{!}{\includegraphics{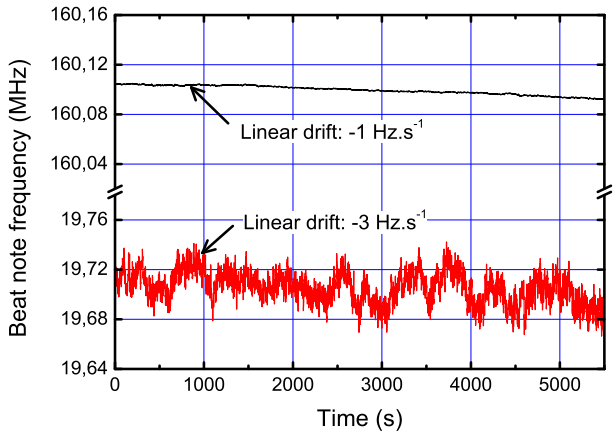}} 
 \caption{(Color online) Red curve: Beat note frequency between LD2 and the OFC. Black curve: beat note frequency between LD1 and LD2.}
 \label{ref_stab}
 \end{figure}

The long-term frequency stability is measured by comparison with the OFC with the method described in Sect.~\ref{rep}. We report on the Fig.~\ref{ref_stab} the beat note frequency between LD2 and the OFC over 5000~s. We show on the same graph the frequency beat note between LD1 and LD2 around 160~MHz at the output of the first mixer. This signal exhibits a slope of -1~Hz.s$^{-1}$ coming from the frequency drift of the VCO on which the beat note is locked. It could be removed by phase-locking the VCO on a stable frequency reference. The frequency stability of LD2 is affected by the combined drifts of LD1 and the VCO. Thus, the beat note between LD2 and the OFC shows a slope of -3~Hz.s$^{-1}$ which is in good agreement with the measurements realized on LD1.

The fractional frequency stability of LD2 compared to the OFC is reported on Fig.~\ref{allan_dev}. It shows a stability under $4 \times 10^{-11}$ from 1~s to 1000~s. The stability of LD2 is better than LD1 from 1~s to 100~s. This difference is due to the fact that the current of LD1 is directly modulated at 100~kHz. This modulation, necessary for the FM spectroscopy technique, degrades the stability of the laser by increasing the high-frequency noise level. Nevertheless, this effect has few impact on LD2 as the frequency noise degradation is above its servo bandwidth.

\section{Conclusion}

To conclude, we proposed and realized an optical setup for rubidium cooling based on commercial off-the-shelf fiber components. We demonstrate the compatibility of each elements for space application. The frequency conversion module, based on pigtailed PPLN waveguide, has been submitted to a set of environmental tests. The successful results of this characterization make it suitable for space applications. A complete experimental laser source has been realized and extensively characterized to ensure the compatibility of its performances with sub-Doppler atomic cooling. A new method has been proposed to guarantee the stability and the robustness of the laser lock on very long time scale using a numeric integrator. Further development could be performed to improve the system and to realize new functions. For instance, adding a fiber phase modulator at the output of the laser diode could enable to generate sidebands for driving Raman transitions~\cite{menoret2011}.

To a great extent, this study shows that this kind of architecture, using commercial off-the-shelf space qualified components, is a relevant solution to promote the use of cold atoms in the frame of scientific space missions. First, no development of specific components or subsystems is needed. Second, the assembly of the source is eased by the use of fiber components. Risks of misalignment are also reduced by limiting the free space propagation of the beams. Finally, the frequency doubling architecture enables to use efficient components at 1560~nm already developed for telecom purposes which confers excellent performances to the laser source. Then, the cost and the risks of industrial development for a laser source flight model could be drastically reduced. Industrial studies of this concept will be performed in the frame of the STE-QUEST space study.

\begin{acknowledgement} We would like to thank AdvEOTec company for environmental characterization of the frequency doubling module. We also thank F.X. Esnault and A. Gauguet for fruitful discussions and careful readings. We thank M. Lours from SYRTE laboratory for providing electronic locking modules.
\end{acknowledgement}

\appendix
\section{Environmental tests}
\label{environmental_tests}

In this part, we give the results of the life tests and environmental tests performed on the wavelength conversion module. The conversion efficiency $\eta$ and the phase-matching temperature $T_{\mathrm{pm}}$ were measured for an input power of $P_{\mathrm{in}}$=20~mW before and after each test. These two parameters are determined with a respective relative uncertainty of 13$\%$ and 5$\%$ at 3$\sigma$. The life tests, reported in  Tab.~\ref{life}, were performed during 2000 hours on three different devices (D1, D2 and D3). The environmental tests, chronologically reported in Tab.~\ref{environmental}, were carried on the device D2.

\begin{table}[!h]
%\centering
\caption{Life tests}
\label{life}
\begin{tabular}{llll}
\hline\noalign{\smallskip}
Device & Conditions & $\eta(\%/\mathrm{W})$ & $T_{\mathrm{pm}}$($^{\circ}$C)  \\
\noalign{\smallskip}\hline\noalign{\smallskip}
D1 & Initial & 254 & 44.3\\
D1 & 2000~h/400~mW & 207 & 44.0\\
D2 & Initial & 217 & 40.4\\
D2 & 2000~h/700~mW & 241 & 41.4\\
D3 & Initial & 270 & 40.9\\
D3 & 2000~h/1000~mW & 270 & 41.1\\
\noalign{\smallskip}\hline
\end{tabular}
% Or use
%\vspace*{5cm}  % with the correct table height
\end{table}

\begin{table}[!h]
%\centering
\caption{Environmental tests on D2}
\label{environmental}
\begin{tabular}{llll}
\hline\noalign{\smallskip}
Test & Conditions & $\eta(\%/\mathrm{W})$ & $T_{\mathrm{pm}}$($^{\circ}$C)  \\
\noalign{\smallskip}\hline\noalign{\smallskip}
- & Initial & 217 & 40.4\\
Life Test & 2000~h/700~mW & 241 & 41.4\\
Gamma & 10~krad & 237 & 40.4\\
Gamma & 50~krad & 241 & 40.3\\
Thermal cycling & -40$^{\circ}$C/+60$^{\circ}$C & 240 & 41.0\\
Vibration & 20~grms & 209 & 40.7\\
Shock & 500~g & 235 & 40.6\\
Thermal cycling & -55$^{\circ}$C/+75$^{\circ}$C & 231 & 42.3\\
\noalign{\smallskip}\hline
\end{tabular}
% Or use
%\vspace*{5cm}  % with the correct table height
\end{table}

\newpage

% BibTeX users please use
% \bibliographystyle{}
% \bibliography{}

\end{document}